\documentclass[10pt,letterpaper]{article}
\usepackage[top=0.85in,left=2.75in,footskip=0.75in]{geometry}

\usepackage{amsmath,amssymb}

\usepackage{changepage}

\usepackage[utf8x]{inputenc}

\usepackage{textcomp,marvosym}

\usepackage{cite}

\usepackage{nameref,hyperref}

\usepackage{microtype}
\DisableLigatures[f]{encoding = *, family = * }

\usepackage[table]{xcolor}

\usepackage{array}

\newcolumntype{+}{!{\vrule width 2pt}}

\newlength\savedwidth

\raggedright
\usepackage[aboveskip=1pt,labelfont=bf,labelsep=period,justification=raggedright,singlelinecheck=off]{caption}

\makeatletter
\renewcommand{\@biblabel}[1]{\quad#1.}
\makeatother

\usepackage{lastpage,fancyhdr,graphicx}
\usepackage{epstopdf}
\fancyheadoffset[L]{2.25in}
\fancyfootoffset[L]{2.25in}
\begin{document}
\vspace*{0.2in}

\begin{flushleft}
{\Large
\textbf\newline{Finding morphology points of electrocardiographic signal waves using wavelet analysis} 
}
\newline
\\
Alena I. Kalyakulina\textsuperscript{1*},
Igor I. Yusipov\textsuperscript{1},
Victor A. Moskalenko\textsuperscript{1},
Alexander V. Nikolskiy\textsuperscript{2},
Artem A. Kozlov\textsuperscript{3},
Nikolay Yu. Zolotykh\textsuperscript{1},
Mikhail V. Ivanchenko\textsuperscript{1}
\\
\bigskip
\textbf{1} Institute of Information Technologies, Mathematics and Mechanics, Lobachevsky State University, Nizhniy
Novgorod, Russia
\\
\textbf{2} Department of Cardiovascular Surgery, City Clinical Hospital No 5, Nizhniy Novgorod, Russia
\\
\textbf{3} Department of Anaesthesiology and Reanimation, Semashko Regional Clinical Hospital, Nizhniy Novgorod, Russia
\\
\bigskip

* alena.kalyakulina@itmm.unn.ru

\end{flushleft}
\section*{Abstract}
A new algorithm has been developed for delineation of significant points of various electrocardiographic signal (ECG) waves, taking into account information from all available leads and providing similar or higher accuracy in comparison with other modern technologies. The test results for the QT database show a sensitivity above 97\% when detecting ECG wave peaks and 96\% for their onsets and offsets, as well as better positive predictive value compared to the previously known algorithms. In contrast to the previously published algorithms, the proposed approach also allows one to determine the morphology of waves. The segmentation mean errors of all significant points are below the tolerances defined by the Committee of General Standards for Electrocardiography (CSE).


\section*{Introduction}\label{introduction}

Wavelet analysis as a special type of linear transformation is widely used in the signal processing, which are nonstationary in time or inhomogeneous in space \cite{Grossman1984, Meyer1993}. The results of this analysis allow us to identify not only the overall frequency response of the signal, but also information about local coordinates, where certain groups of frequency components of the signal are detected or changed. Therefore this method finds its application in many problems of radiophysics \cite{Astafieva1996, Dremin2001}, theory of synchronization \cite{Postnikov2009}, acoustics \cite{Anisimov2008}, neurodynamics and neurophysiology \cite{Koronovskiy2013}, recognition and classification of images \cite{Dyachenko2005}. In addition, the analysis of wavelets is used in solving problems of the processing of electrocardiograms signal \cite{Teptin2011}, including delineation of specific waves and complexes.

The electrocardiogram (ECG) is a recording of the electrical activity of the heart, obtained by the electrodes, located on the human body. This is one of the most powerful diagnostic tools for analyzing cardiac activity and the most frequently used noninvasive test of primary healthcare to determine the heart rate and rhythm disturbances \cite{Hooper2001, Fairweather2007}.

Ambulant ECG analysis requires processing of signals subject to various disturbances \cite{Lee2007}. They include power interference, electrode contact noise, motion and muscle artifacts, baseline drift and high-frequency noises. Since such interference can overlap the cardiac component in the space-time and frequency domains, the isolation of a weak cardiac component from the noisy ECG signal is a rather complex problem. Radiophysical signal processing techniques proves themselves in solving this type of problem \cite{Ivlev2004}. The noise reduction of ECG signals facilitates its processing and helps to obtain the most accurate information about the heart work.

\begin{figure}[!h]
	\includegraphics*[width=80 mm]{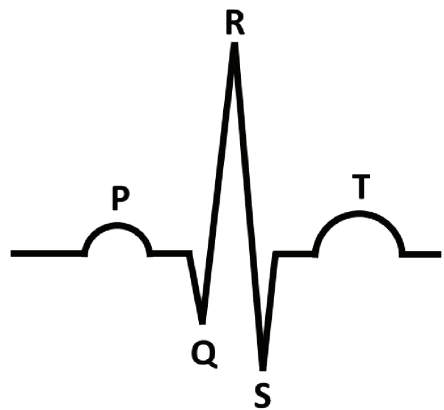}\\
	\caption{{\bf ECG signal.}
		Schematic representation of the main complexes and waves of the ECG signal.}
	\label{fig1}
\end{figure}

\begin{figure}[!h]
	\includegraphics*[width=120 mm]{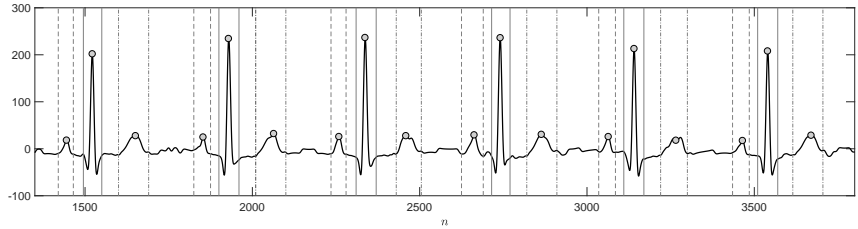}\\
	\caption{{\bf Example of ECG delineation.}
		The symbols show peaks of QRS, P, and T complexes, lines represent complexes boundaries (solid lines are the QRS complex, the dashed lines are the P wave, the dot-dash lines are the T wave).}
	\label{fig2}
\end{figure}

On the electrocardiogram, the following waves and complexes are identified: QRS complex, P and T waves  (Fig. \ref{fig1}). The P wave on the ECG reflects the process of depolarization of the right and left atriums, the QRS complex corresponds to the process of propagation of excitation on the ventricular myocardium, and wave T shows the process of rapid terminal repolarization of the ventricular myocardium. Analysis of their amplitudes, shapes (morphologies) and durations allows to detect cardiac rhythm disorders and cardiovascular diseases, such as ischemia and myocardial infarction \cite{Khan2009}. In addition to the QRS complex, for diagnostics it is also necessary to find waves P and T, because there are diseases with the pathological changes of these waves. For example, myocardial or atrial hypertrophy changes the form of the P wave, and ischemia changes the shape of the T wave. The problem of ECG delineation consists of determining the amplitudes and time intervals for the location of the waves and complexes. An automation of the solution of this problem, both at the hardware level and in the computer processing of the ECG signals, is of extreme interest now (Fig. \ref{fig2}).

Accurate performance of the delineation is a complicated problem due to a number of reasons. For example, P wave has small amplitude and can be hidden due to a movement of the electrodes or some muscle noise. P and T waves can be biphasic, which complicates the precise determination of their onsets and offsets. Moreover, some cardiac cycles may not contain all standard ECG waves, for example, there may be no P wave, and with an increased heart rate, it may be partially overlapped  with the T wave of the previous beat.

The first stage of the ECG signal delineation consists of the QRS complex detection, which in most cases is the most amplitude wave of the cardiac cycle. In subsequent processing P wave, QRS complex and T wave reference points (onset, peak, offset) are detected. The cyclic nature of the ECG signal and its spectral components make the ECG a suitable candidate for wavelet analysis. Methods, based on wavelet transforms, were proposed by many authors \cite{Martinez2004, Addison2005, DiMarco2011, Bote2017} and were based on the well-known method proposed by Li \cite{Li1995}.

\begin{figure}[!h]
	\includegraphics*[width=120 mm]{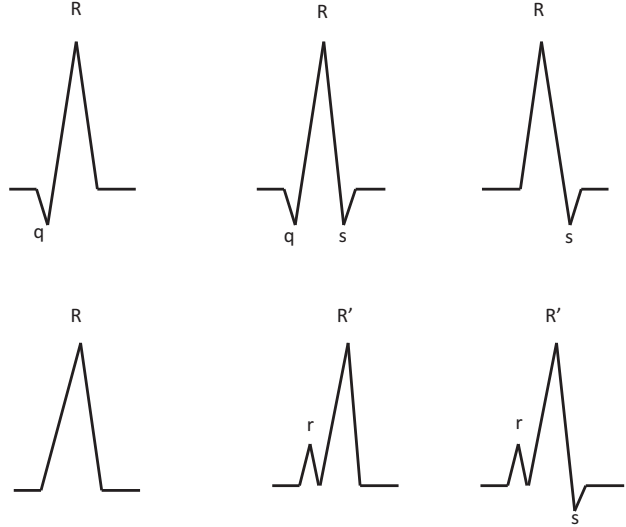}\\
	\caption{{\bf Example of QRS complex forms.}
		An example of the most common types of QRS complex forms.}
	\label{fig3}
\end{figure}

Unfortunately, the cited algorithms detect only the main wave points of the ECG signal: onsets, peaks and offsets. This information may not be enough to establish a diagnosis. Thus, when analyzing the ECG signal to detect symptoms of cardiovascular diseases, it is important to know not only the time marking of the waves, but also their shape (morphology), namely, the presence of several peak phases, additional wave points, the location and orientation of the Q and S peaks of the QRS complex etc (examples of forms are shown on Fig. \ref{fig3}). Such complete information about the markup of all ECG signal wave points, their mutual arrangement and directions, the intervals between significant points can be used by a physician to determine symptoms of cardiac rhythm disturbances and the presence of cardiovascular diseases.

In this paper we propose a new algorithm for determining significant points of ECG waves, taking into account all available leads and providing similar or higher accuracy in comparison with known technologies. On the QT Database our algorithm gives sensitivity above $97\%$ for ECG wave peaks and $96\%$ for their onsets and offsets, as well as better positive predictive value (PPV) than in previously known solutions. The main difference from the known approaches is that  the proposed approach also allows to determine the morphology of the waves. The segmentation mean errors of all significant points are below the tolerances defined by the Committee of General Standards for Electrocardiography (CSE).

\section{Wavelet transform}\label{wavelet}

The wavelet transform is a decomposition of a signal into a combination of a set of basis functions obtained by dilation 
(with the scaling factor $a$) and translation (with the shift parameter $b$) of one mother wavelet. 
Thus, the wavelet transform of a signal $x(t)$ is defined as follows:

\begin{equation}
	\label{eq1}
	W_a x\left(b\right) = \frac{1}{\sqrt{a}}\int_{-\infty}^{+\infty}x\left(t\right)\psi\left(\frac{t-b}{a}\right) dt, \quad a>0.
\end{equation}

The larger the scaling factor $a$, the wider the base function, and therefore the corresponding coefficient gives information on the lower frequency components of the signal and vice versa. If the mother wavelet $\psi(t)$ is a derivative of the some smoothing function $\theta(t)$, then the wavelet transform of the signal $x(t)$ on the scale $a$ has the form:

\begin{equation}
	\label{eq2}
	W_a x\left(b\right) = -a\, \frac{d}{db}\int_{-\infty}^{+\infty}x\left(t\right)\theta_a\left(t-b\right) dt, 
\end{equation}
where $\theta_a\left(t\right)=\left(1/\sqrt{a}\right)\theta\left(t/a\right)$ is a scaled version of the smoothing function.

The wavelet transform on the scale $a$ is proportional to the derivative of the filtered version of the signal with a smoothing impulse response on the $a$ scale. Therefore, the zeros of the wavelet correspond to local maxima or minima of the smoothed signal on different scales, and the maximal absolute values ​​of the wavelet transform correspond to the maximum slopes of the filtered signal.

The scaling factor $a$ and the shift parameter $b$ can be discretized. Usually a binary grid is considered: $a = 2^k$ and $ b = 2^k l$. In this case, the basic functions are represented in the following form:

\begin{equation}
	\label{eq3}
	\psi_{k,l}\left(t\right)=2^{-k/2}\psi\left(2^{-k}t-l\right) \qquad (k,l=1,2,\dots).
\end{equation}

A discrete wavelet transform of the signal $x$ is obtained by applying a set of filters. 
First, the signal is passed through a low-frequency filter with an impulse response $G$, and the convolution is obtained:

\begin{equation}
	\label{eq4}
	y\left[n\right] = \sum \limits _{{k=-\infty }}^{\infty }{x[k]G[n-k]}.
\end{equation}

Simultaneously, the signal is decomposed using the high-pass filter $H$ in the same way:

\begin{equation}
	\label{eq5}
	y\left[n\right] = \sum \limits _{{k=-\infty }}^{\infty }{x[k]H[n-k]}.
\end{equation}

After applying the high-pass filter $G$, the so-called detailed coefficients are obtained, after applying the low-pass filter $H$, approximation coefficients are obtained. These two filters are called quadrature mirror filters.

\begin{figure}[!h]
	\includegraphics*[width=120 mm]{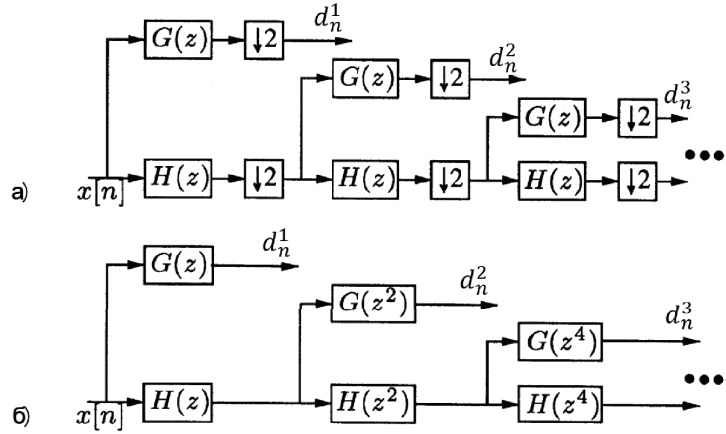}\\
	\caption{{\bf Two types of the filter bank for a discrete wavelet transform.}
		(top) Mall's algorithm. (bottom) Algorithm \'{a} trous.}
	\label{fig4}
\end{figure}

For time-sampled signals, the discrete wavelet transform, according to Mall's algorithm, reduces to constructing a bank of high and low frequency filters \cite{Mallat1989}. The high and low frequency filters are applied to the source signal. This gives detailed and approximation coefficients of the first scale, respectively. The application of high and low frequency filters to the approximation coefficients of the first scale gives the detailed and approximation coefficients of the second scale. This process can be continued. The standard algorithm involves subsampling of the coefficients, as shown in Fig.\,\ref{fig4}\,(top).

In the problem of ECG signal delineation there is no need for subsampling, because the time resolution of wavelet coefficients on large scales $a$ decreases. Therefore, the algorithm \textit{\`{a} trous} \cite{Cohen1996} is used, in which the sampling rate on all scales is the same and there is no subsampling; see Fig.\,\ref{fig4}\,(bottom).

\begin{figure}[!h]
	\includegraphics*[width=120 mm]{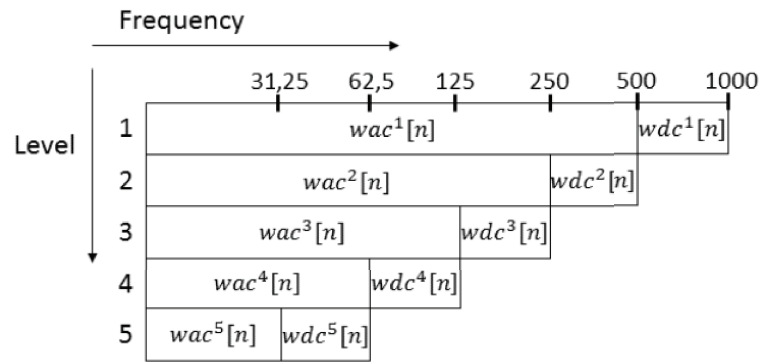}\\
	\caption{{\bf Detailed coefficients of the discrete wavelet transform.}
		Detailed coefficients characterize the different frequency ranges for the scales $ k = 1, 2, 3, 4, 5 $ at a signal sampling frequency of 1000 Hz. \textit{wac} means \textgravedbl wavelet approximation coefficients\textacutedbl, \textit{wdc} means \textgravedbl wavelet detailed coefficients\textacutedbl.}
	\label{fig5}
\end{figure}

In this paper we use the biorthogonal wavelets $1.5$ proposed in \cite{Mallat1992}. The filter coefficients of high ($G[k]$) and low ($H[k]$) frequencies filters are presented in Table \ref{table1}. The parameters of the delineation algorithm depend significantly on the choice of the mother wavelet. The frequency characteristics of the first five scales for the detailed coefficients are presented in the Fig.\,\ref{fig5}. When the sampling frequency of the signal is $1000$~Hz, the detailed coefficients on the first scale characterize the frequency range $500$--$1000$~Hz; the detailed coefficients of the second scale, $250$--$500$~Hz; the detailed coefficients of the third scale, $125$--$250$~Hz; and so on. The cumulative consideration of the detailed coefficients of different scales allows us to find not only the reference points of the ECG wave signals (onset, peak, offset), but also to consider in details the morphology of each detected complex.

\begin{table}[!ht]
	\begin{adjustwidth}{-2.25in}{0in} 
		\centering
		\caption{
			{\bf Table of high and low frequency filter coefficients corresponding to biorthogonal wavelets 1.5.}}
		\begin{tabular}{|l|l|}
			\hline
			&\\[-1em]
			$H[0]=0.0165728152$ & $G[0]=0$ \\
			\hline
			&\\[-1em]
			$H[1]=-0.0165728152$ & $G[1]=0$ \\
			\hline
			&\\[-1em]
			$H[2]=-0.1215339780$ & $G[2]=0$ \\
			\hline
			&\\[-1em]
			$H[3]=0.1215339780$ & $G[3]=0$ \\
			\hline
			&\\[-1em]
			$H[4]=0.7071067812$ & $G[4]=-0.7071067812$ \\
			\hline
			&\\[-1em]
			$H[5]=0.7071067812$ & $G[5]=0.7071067812$ \\
			\hline
			&\\[-1em]
			$H[6]=0.1215339780$ & $G[6]=0$ \\
			\hline
			&\\[-1em]
			$H[7]=-0.1215339780$ & $G[7]=0$ \\
			\hline
			&\\[-1em]
			$H[8]=-0.0165728152$ & $G[8]=0$ \\
			\hline
			&\\[-1em]
			$H[9]=0.0165728152$ & $G[9]=0$ \\
			\hline
		\end{tabular}
		\label{table1}
	\end{adjustwidth}
\end{table}

\section{ECG delineation}\label{delineation}

In this section we describe our ECG delineation algorithm.

\subsection{General information}\label{delineation:common}

An electrocardiogram is a set of signals recorded from different leads. Discrete wavelet transform is applied to each lead of the filtered ECG signal without subsampling. High-scaled detailed coefficients are used to find the peaks of QRS complexes, P and T waves and the approximate location of their boundaries. The detailed coefficients of the low scales are used to refine the boundaries of the complexes, and also to analyze their morphologies.

During the algorithm, QRS complexes are first delineated (as a rule, they have the largest amplitude) on each lead separately. Then, based on the results of the delineation of all leads, the incorrectly found complexes are removed, the missing complexes are added, and the boundaries are corrected. After delineation of QRS complexes, the wave T, then the wave P (as a rule, T wave has greater amplitude than P wave) is delineated in the same way.

The process of delineation has the following general steps:

\begin{enumerate}
	\item[(1)] peak finding;
	\item[(2)] defining of primary boundaries;
	\item[(3)] clarification of boundaries, definition of morphology;
	\item[(4)] refinement, based on delineation results for all leads.
\end{enumerate}

Steps (1) and (2) are traditional for delineation algorithms using wavelet transform and described in \cite{DiMarco2011, Martinez2004}.
Steps (3) and (4) are the main innovations of the paper. In addition, we propose a number of changes in steps (1) and (2).

\begin{figure}[!h]
	\includegraphics*[width=120 mm]{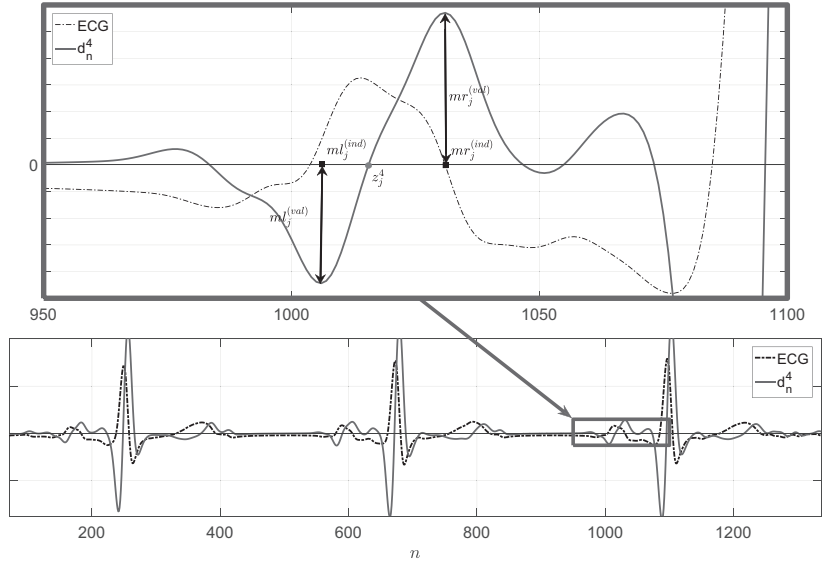}\\
	\caption{{\bf Example of considered ECG elements.}
		Example of ECG (solid line) with the detailed coefficients of the $4$th scale (dash-dot line). The enlarged image illustrates the notation from \ref{delineation:common}. The circle $z_j^ 4$ corresponds to the considered zero crossing. The squares $ml_j^{(ind)}$ and $mr_j^{(ind)}$ are the indexes of the left and right extrema, respectively.}
	\label{fig6}
\end{figure}

The main subject of the delineation algorithm is the detailed coefficients, denoted as $d_n^k$, where $k$ is the index of scale, and $n$ is the index of the signal measurement. On a given scale $k$, all zeros $z_j^k $ (where $j$ is the number of zero crossing) of the detailed coefficients are found. Since the detailed coefficients are proportional to the derivative of the filtered signal, where the scale $k$ is responsible for the degree of filtration, the found zeros $z_j^k$ correspond to the extrema of the original signal. The most amplitude extrema are the peaks of the main waves and complexes of the ECG signal. For each zero, there are the following characteristics, presented in Fig.\,\ref{fig6}:

\begin{enumerate}
	\item $ml_j$ is the global extremum located to the left of $z_j^k$ (the left boundary of the extremum search is limited to the previous zero crossing or the beginning of the signal). It is characterized by the index and the value ($ml_j^{(ind)}$ and $ml_j^{(val)}$, respectively).
	\item $mr_j$ is the global extremum located to the right of $z_j^k$ (the right boundary of the extremum search is limited to the next zero crossing or the end of the signal). It is characterized by the index and value ($mr_j^{(ind)}$ and $mr_j^{(val)}$, respectively).
	\item $a_j$ is the amplitude of zero crossing, which is determined by the values of the extremes on both sides of it: $a_j = |ml_j^{(val)}| + |mr_j^{(val)}|$.
\end{enumerate}

Also in the delineation algorithm, it is necessary to find local extrema of the detailed coefficients:
\begin{enumerate}
	\item[(1)] coordinate extrema, i.e. positive maxima and negative minima;
	\item[(2)] incoordinate extrema, i.e. positive minima and negative maxima.
\end{enumerate}

\subsection{QRS delineation}\label{delineation:qrs}

As a rule, QRS complex has the largest signal amplitude. Therefore zero crossings $z^k$ with the maximum amplitude $a$ correspond to the one of the waves of this complex. However, QRS complex can be complicated and in addition to the standard points Q, R and S may have additional peaks of R. Also, for some cardiovascular diseases, the height of the T wave can be comparable to the height of R \cite{Khan2009}, which complicates the delineation.

\subsubsection{QRS peak}\label{delineation:qrs:peak}

To determine the peak of the complex QRS, we use detailed coefficients corresponding to frequencies from $30$ to $100$~Hz.
If the original signal sampling frequency is $500$~Hz, then it is necessary to consider the detailed coefficients with the scale numbers $k = 3$ or $k = 4$ (an example of the detailed coefficients for $k = 4$ with a $500$~Hz signal frequency is shown in Fig.\,\ref{fig6}).

We search for all of zero crossings $z^k$. If a zero crossing $z_j^k$ is found, it is stored as the current candidate for the QRS complex peak. If a zero crossing $z_i^k$ with larger amplitude (i.e. $a_i>a_j$) is found in the window of $250$~ms after the current candidate, then $z_i^k$ replaces the zero crossing $z_j^k$ as a candidate. Similar search is performed for $z_i^k$ in the $250$~ms window. This process stops, when there are no new suitable candidates in the search window. Then we form a list of primary candidates $\overline{z}^k$ for the role of the QRS complex peak. The choice of the time window ($250$ ms) corresponds to the maximum possible expected heart rate, $240$~beats per minute.

Then we check the candidates and consider a part of the ECG signal ($w^{tr}$), which definitely contains QRS complexes. For example, if the signal is long enough, we take $8$ seconds, during which there are guaranteed to have at least $4$ cardiac beats (QRS complexes). It corresponds to the lowest possible heart rate of $30$ beats per minute. To make a decision about the correctness of the candidate $\overline{z}_j^k$, it is necessary to calculate an additional characteristic, the zero crossing amplitude in a certain window $\Delta$ ($a_j^\Delta$), which is defined as the difference between maximal and minimal values of the detailed coefficients $d_n^k$ in the interval $\left[\overline{z}_j^k-\Delta, \overline{z}_j^k+\Delta\right]$. The value of $\Delta$ is $100$~ms and it corresponds to the typical duration of the QRS complexes. In the training window $w^{tr}$, the following value is calculated:

\begin{equation}
	\label{eq6}
	\varepsilon^k = \frac{p_1^{QRS}}{N} \sum_{\overline{z}_j^k in w^{tr}} a_j^\Delta,
\end{equation}
where $p_1^{QRS}$ is a numerical parameter (whose optimal value is presented in Table \ref{table2}, $N$ is the number of guaranteed peaks of QRS complexes in the given time interval $w^{tr}$ (with taking into account the minimal possible heart rate), and the sum in the formula (\ref{eq6}) is taken over $N$ largest values, which are found in the given interval.

For the next zero crossing $\overline{z}_j^k$ (primary candidate), which is outside the training window $w^{tr}$, the condition $a_j^\Delta > \varepsilon^k$ is checked. If the condition is satisfied, then the current zero crossing is finally defined as the peak of the QRS complex. After this, the value $\varepsilon^k$ is recalculated, taking into account the new detected peak: the number of guaranteed peaks of QRS complexes includes current zero crossing, and its value $a_j^\Delta$ is added to the sum on the right side of (\ref{eq6}), while the first element of this sum is excluded from consideration ($N$ remains unchanged). Due to this approach, delineation adaptation is achieved in case of local ECG signal changes. It is also possible to restore delineation in the training window $w^{tr}$, if this approach is applied from right to left, after finding all peaks of the QRS complexes outside the training window.

\subsubsection{QRS primary boundaries}\label{delineation:qrs:borders}

The onset of the QRS complex is sought to the left of $ml_j^{(ind)}$, where $j$ is the number of the zero crossings corresponding to the peak of this complex. In a fixed window with the length of $150$~ms we search for all local coordinate extrema. Then from the right to the left we look over all local extrema with modules greater than $t_{on}=p_2^{QRS}\cdot a_j $, where $p_2^{QRS}$ is a numerical parameter whose value is given in Table~\ref{table2}. The index of the last extremum, whose absolute value exceeded $t_{on}$, will be the starting point of the search for QRS onset. We store the value of this extremum, which is denoted as $\overline{m}$. To avoid incorrect delineation of complex morphologies of the QRS complex, we need to perform additional verification. If in the previous paragraph an extremum with an absolute value less than $t_{on}$ was found, then the following extrema should be checked: their absolute values should not exceed $T_{on}=p_3^{QRS}\cdot a_j $ (value $p_3^{QRS}$ is shown in Table~\ref{table2}. If this condition is violated (for example, when the QRS complex has a complicated M-shaped morphology), then the index of the last local extremum, whose module exceeded the value $T_{on}$ (value $\overline{m}$), will also be updated.

We start with the beginning index and the value of $m$, defined in the previous paragraph, move to the left and find two indexes:

\begin{enumerate}
	\item(1) index of threshold value $b_{on}=p_4^{QRS}\cdot \overline{m}$ crossing;
	\item(2) index of an incoordinate extremum, if the latter exists before the next zero crossing.
\end{enumerate}
The maximum of this two indexes will be the index of the QRS onset.

Searching for the QRS offset is doing in the same way in the opposite direction. Instead of $t_{on}$ and $b_{on}$, the values $t_{off}=p_5^{QRS}\cdot a_j$ and $b_{off}=p_6^{QRS}\cdot \overline{m}$ is considered (Table \ref{table2}).

\subsubsection{QRS boundary refinement and morphology determination}\label{delineation:qrs:morphology}

For ECG signal delineation, the authors of \cite{DiMarco2011, Martinez2004} use detailed coefficients of only high scales. One of the main innovations of our paper is using the detailed coefficients of low scales, which characterize the ECG signal components with higher frequencies. This allows to form the morphology of each certain complex in the form of a list of points. Each point of this morphology is characterized by the name, index and value of the potential. In addition, this approach allows to analyze such complex phenomena as late ventricular and atrial potentials \cite{Latfullin2009}.

To determine the morphology of the complex, it is necessary to consider not only large scales of the detailed coefficients, but also the detailed coefficients of low scales that characterize the high-frequency components of the signal. For example, when the original sample rate is $500$~Hz, the scale $k^*=1$ is considered. First, we form a list of all zero crossings of the detailed coefficients at large scales $d_n^k$, which are inside the primary boundaries of the complex. The zero crossing corresponding to R peak is known. If current zero crossing is not the first and not the last in the given boundaries, then the first zero crossing corresponds to Q peak of the complex, and the last one to S peak. If R peak is the first zero crossing, then the QRS complex has no Q peak, if it is the last one, then there is no S peak. If a single zero crossing is found, it means, that the current QRS complex does not have both Q peak and S peak. Then, in the primary boundaries, we find all zero crossings of the low-scale $k^*$ detailed coefficients. Among these zeros crossings we need to find a correspondence between the peaks Q, R and S. If there is zero crossings at the scale $k^*$ between Q and R (or between R and S), the corresponding peaks are recorded in the morphology of the QRS complex.

The next step is to refine the boundaries of the complex. We need in correction of the complex onset only when Q peak on the $k$ scale is detected and the offset correction performs only in case of S peak detection. If both Q and S peaks were not detected at the previous stage, then we use indexes obtained after the \ref{delineation:qrs:borders} step. Otherwise, for each previously delineated complex, we calculate its duration $L^{QRS}$ and the deviation from the maximum allowable value of the QRS complex duration ($p_7^{QRS}$): $\eta^{QRS}=p_7^{QRS}-L^{QRS}$. If $\eta^{QRS}<0$, then there is no need in boundaries refinement (the result obtained after the stage \ref{delineation:qrs:borders} is stored). Otherwise, when correcting the boundaries of the complex, they can be extended by $\eta^{QRS}/2$ on the left when finding the onset of the complex and can be extended by the same value on the right when finding the offset of the complex. In the interval $\left[on-\eta^{QRS}/2,on\right]$ we find zero crossings and local extrema on the scale $k^*$ of the detailed coefficients ($on$ is the index of the Q peak). If zero crossings were found, then $zc_{on}$ with the maximal value $|mr_{on}^{(val)}|$ is selected as the onset index of the QRS complex. If no zero crossings were found, then the index of the last incoordinate extremum in the interval $\left[on-\eta^{QRS}/2, on\right]$ is the onset of the QRS. If the latter were not found, then the index found in the paragraph \ref{delineation:qrs:borders} is selected as the onset. Refinement of the QRS complex offset is carried out in a similar way.

\subsubsection{QRS complexes refinement based on delineation of all leads}\label{delineation:qrs:multilead}

This modification is used for ECG signals with a large number of leads. After the previous stages of delineation, the following situations are possible:

\begin{enumerate}
	\item[(1)] there is no delineation of the complex on a current lead;
	\item[(2)] there are mistakes of delineation (for example, the high-amplitude T wave was incorrectly recognized by the algorithm as a QRS complex);
	\item[(3)] boundaries of the complex are incorrect (for example, when the QRS onset intersects with P wave).
\end{enumerate}

\begin{figure}[!h]
	\includegraphics*[width=120 mm]{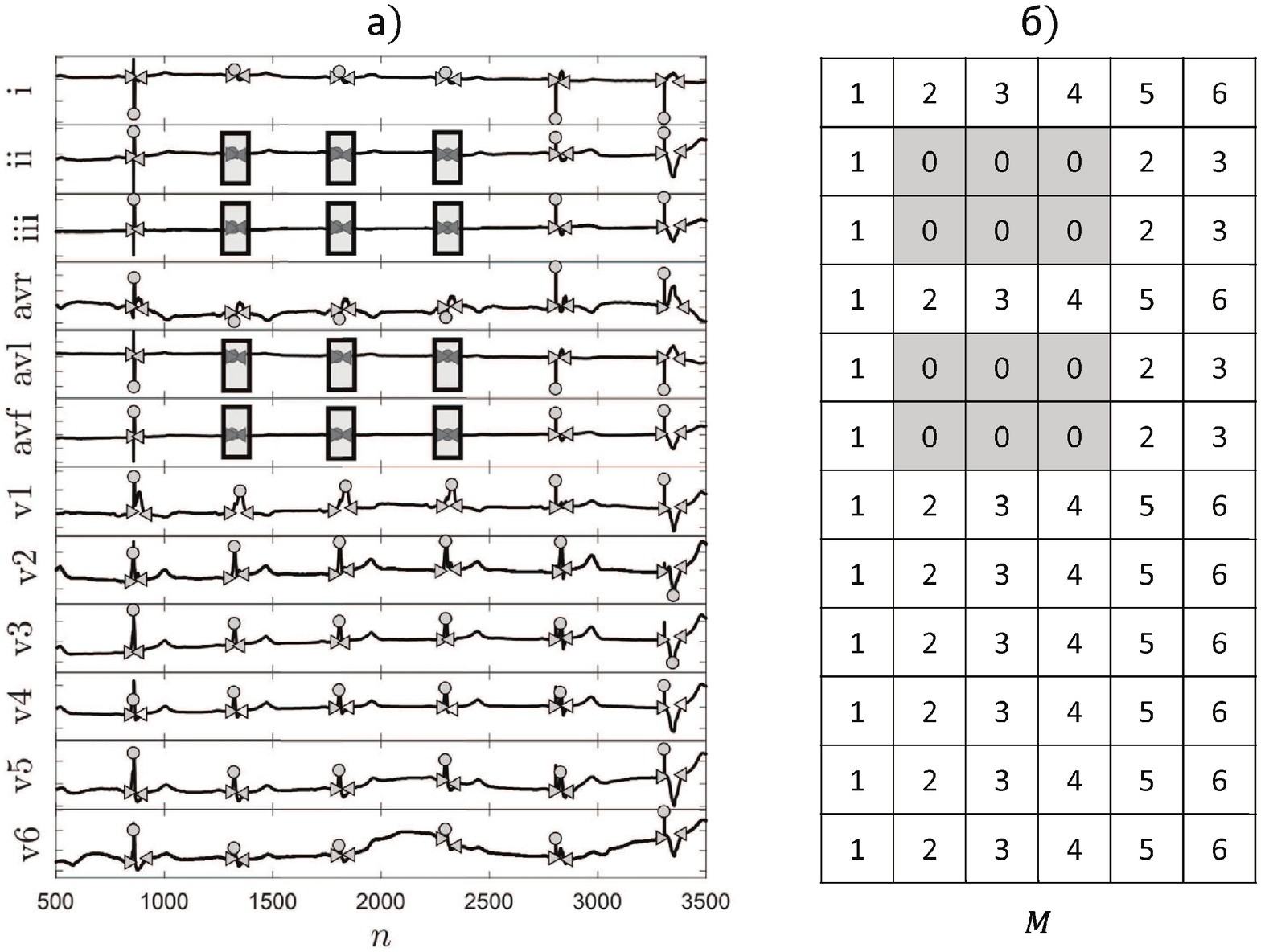}\\
	\caption{{\bf 12-lead ECG example.}
		(left) Example of a 12-channel delineated ECG with a pacemaker. The boundaries of the complexes are shown by triangles, peaks -- by circles. After the first stages of delineation (\ref{delineation:qrs:peak}, \ref{delineation:qrs:borders}, \ref{delineation:qrs:morphology}, there are complexes, marked by rectangles at the part of the leads. Due to the refinement taking into account the set of leads, their delineation was restored. (right) Matrix of the correspondence $M$ for the case, shown in the Fig. \ref{fig7}(left), before refinement for the set of leads. Zeros in $M$ correspond to missing non-delineated complexes.}
	\label{fig7}
\end{figure}

The refinement taking into account a set of leads avoids most of the above disadvantages of the basic version of the algorithm, based on \cite{DiMarco2011, Martinez2004}. In addition to general improvement in the quality of complexes delineation, this approach allows to solve delineation problems in complicated cases, for example, when there is a pacemaker and some QRS complexes have a significantly larger amplitude than others; see Fig.\,\ref{fig7}\,(left). In this case some complexes remain undelineated after the first stages; in Fig.\,\ref{fig7}\,(left) they are marked by rectangles. But the refinement taking into account the set of leads allows to restore their delineation.

First, based on the results of delineation, the matching matrix $M$ is constructed. The rows of this matrix correspond to individual leads (an example of the matrix is shown in Fig.\,\ref{fig7}(right). The elements of the matrix are numbers of complexes for a current lead, or $0$, which means that the given complex is not found on the current lead. Before the matrix construction, the mean length of QRS complex $\langle QRS \rangle$ is computed on the entire ECG. At the beginning of the process of constructing the matrix $M$ the lead with the maximal number of delineated complexes is selected and a list $L$ of all found complexes is formed. This list stores the average values of the onset, peak and offset indexes, as well as their standard deviations. Further, for each lead, for each delineated complex, its correspondence in $L$ is found by calculating the minimal difference of the complex peak indexes. The resulting sequence number is stored into the matrix $M$. If the minimal difference exceeds the value $p_8^{QRS} \cdot \langle QRS \rangle$, then a new complex is stored in the list $L$ (the value $p_8^{QRS}$ is presented in Table~\ref{table2}). After forming the matrix $M$ (see Fig.\,\ref{fig7}\,(right)) each its column contains in the corresponding lead the number of complex or $0$ if the given complex was not found.

If a column of $M$ contains zeros, which amount not exceeds $1/3$ of the total number of all leads, then in the corresponding leads, we need to add delineation of the complexes, missed at the previous stages. The primary boundaries of the complex are the averaged values of the boundaries of the same complex from the list $L$. As a peak, zero crossing with the maximal amplitude within these boundaries is taken. After that, the approaches described in \ref{delineation:qrs:morphology} are applied to the new complex: the morphology is restored and the borders are refined.

If a column of $M$ contains non-zero elements, which amount not exceeds $1/3$ of the total number of all leads, we remove incorrect delineation in the corresponding leads.

Also, at this stage we perform a refinement for complexes with incorrect boundaries. If the peak index of the current complex deviates from the average value in $L$ ($m_{peak}$) by the value greater than $p_9^{QRS} \cdot \sigma_{peak}$, where $\sigma_{peak}$ is the standard deviation for the peak of the considered complex, then the new delineation is performed, taking into account the averaged boundary values.

\subsection{T delineation}\label{delineation:t}

The next stage is the T wave delineation, which follows the QRS complex and, as a rule, has a larger amplitude $a$ than P wave.

\subsubsection{T peak detection}\label{delineation:t:peak}

First, we calculate the length of the RR-interval. To find the peak of the complex, it is necessary to find all zero crossings $z^k$ of the detailed coefficients on the scale $k=3$ (when the original sampling frequency of the signal is 500 Hz) in the following window:

\begin{equation}
	\label{eq7}
	W_t: \left[qrs_{off}+80 \text{ ms}, qrs_{off}+p_1^T\cdot RR\right],
\end{equation}
where $qrs_{off}$ is the offset index of the previous QRS complex, the window of $80$~ms corresponds to the characteristic length of the ST interval, $p_1^T$ is the numerical parameter characterizing the admissible search window for the T wave (given in Table~\ref{table2}). Among all zero crossings, we choose zero crossing $z_j^k$ with the maximal amplitude $a_j$. Then we check the following condition:

\begin{equation}
	\label{eq8}
	a_j>p_2^T\cdot A,
\end{equation}
where $A$ is the averaged value of the amplitudes of the zero crossings, corresponding to the two R peaks (left and right peaks of the delineated T wave). If (\ref{eq8}) is not satisfied, the delineation is terminated and we make a conclusion, that there is no T wave. Otherwise, the index of this zero crossing will be the peak of T wave.

T wave can be biphasic: it has $2$ peaks, each of which corresponds to a separate zero crossing. To check this morphology, a zero crossing $z_{j-1}^k$ is found to the left of the detected peak and two conditions are checked:

\begin{equation}	
	\label{eq9}
	\begin{array}{rcl}
	a_{j-1} & > & p_3^T\cdot a_j, \\
	z_j^k-z_{j-1}^k & < & p_4^T\cdot RR.
	\end{array}
\end{equation}
The first condition guarantees that the detected second T wave has a comparable amplitude, and the second condition imposes a restriction on the distance between the peaks; the values of $p_3^T$ and $p_4^T$ are given in Table~\ref{table2}. If the second peak is not detected on the left, then symmetric conditions are checked for zero crossing $z_{j + 1}^k$ on the right.

\subsubsection{T primary boundaries}\label{delineation:t:borders}

Searching for the T wave onset is carried out from right to left. The starting index of the search is the index of the left extremum of zero crossing $ml_T^{(ind)}$, which corresponds to the peak (the left peak, if the T wave is biphasic). The final search index corresponds to the value $qrs_{off}+80 \text{ms}$, where $qrs_{off}$ is the offset index of the previous QRS complex. To the left of the starting value, we are searching for the index of the threshold value crossing $p_5^T \cdot ml_T^{(ind)}$ by the detailed coefficients $d_n^k$ and the index of the first encountered incoordinate extremum. The maximum of this two indexes is selected as the onset. The T wave offset is determined symmetrically with respect to the peak T: searching for candidate indexes is carried out from left to right (the right search boundary is $qrs_{off} + p_1^T \cdot RR$) and the minimal value is chosen.

\subsubsection{T boundaries refinement and morphology determination}\label{delineation:t:morphology}

As in the case of the QRS complex delineation, low scales of the detailed coefficients characterize the high-frequency components of the ECG signal and they are used to refine the boundaries and restore the morphology of the T wave. When the original sampling frequency of the signal is 500 Hz, the scale $k^*=1$ is considered.

Among all zero crossings of the detailed coefficients on the scale $k^*$ in the $\Delta_T^1$-neighborhood of the initially defined peak, we choose a zero crossing with a maximal amplitude ($\Delta_T^1 = p_6^T \cdot L_T$, where $L_T$ is the current length of the T wave). The index of this zero crossing is stored as the updated T wave peak. A similar procedure applies to the second peak in the case of biphasic T wave. All the remaining zero crossings are also recorded in the morphology. The boundaries are corrected as follows: in the $\Delta_T^2$-neighborhood of the initially defined boundaries, we search for zero crossing sand incoordinate extrema on the scale $k^*$ ($\Delta_T^2 = p_7^T \cdot L_T$). The candidates closest to the primary boundaries are marked as new boundaries of the T wave.

\subsubsection{T waves refinement based on the delineation of all leads}\label{delineation:t:multilead}

This stage is performed similarly to the QRS complexes refinement, based on the results from all leads. We form the matrix of the correspondence $M$ for the waves T and the list of all the detected waves $L$. We also calculate the average duration of the wave $\langle T \rangle$ for the entire ECG signal. Analogously to the parameter $p_8^{QRS}$, the parameter $p_8^T$ is introduced, which is used for the list $L$ formation. Adding and removing complexes is also carried out on the principle of majority: if a complex has not been detected by less than a third of the total number of leads, the delineation is restored on the missed leads using the averaged values of boundaries from $L$. If the complex was found only on a third (or less) of all leads, the delineation is removed. Also, for the T wave, a refinement of waves with mistakes is provided, if the peak index of the current complex deviates from the average value in $L$ ($m_{peak}$) by the value, greater than $p_9^T \cdot \sigma_{peak}$. In this case a new delineation is performed taking into account the average values ​​of the boundaries.

\subsection{P delineation}\label{delineation:p}

The P wave delineation is performed at the very last turn. This wave has the smallest amplitude, therefore, this problem is especially complicated. The complexity of solving the problem of the P wave delineation is generally recognized: this is evidenced by a systematic decreasing of the delineation accuracy of P wave, relative to other ECG signal complexes \cite{DiMarco2011, Martinez2004}. In this paper, we propose to localize the search area, limiting it on both sides to already delineated complexes, which helps to improve the quality of delineation.

The algorithm of P wave delineation practically does not differ from the algorithm for T wave. The main differences are the search boundaries and numerical values of parameters.

First, we calculate the length of the RR-interval, where the P wave is searched. To find the peak of the complex, it is necessary to find all the zero crossings $z^k$ of the detailed coefficients of the scale $k=3$ (when the original sampling rate of the signal is $500$~Hz) in the following window:

\begin{equation}
	\label{eq10}
	W_p: \left[qrs_{on}-p_1^P \cdot RR, qrs_{off}-10 \text{ ms}\right],
\end{equation}
where $qrs_{on}$ is the index of the next QRS complex offset point, $p_1^P$ is a numerical parameter that characterizes the admissible search window for the P wave (shown in Table~\ref{table2}). As a candidate for the P peak, among all zero crossings, we choose the zero crossing $z_j^k$ with maximal amplitude $a_j$, if the following condition is satisfied:

\begin{equation}
	\label{eq11}
	a_j>p_2^P\cdot A,
\end{equation}
where $A$ is the averaged value of the amplitudes of zero crossings, corresponding to the two R peaks (to the left and to the right of the delineated T wave). If (\ref{eq11}) is not satisfied, then P wave is absent. For the P wave, morphology can also exist. To check this morphology on the left and right of the peak, conditions, similar to (\ref{eq9}) with the corresponding parameters $p_3^P$ and $p_4^P$ from the Table \ref{table2} are checked.

The search for the P wave onset and offset is carried out as for the T wave, but with differ values of the numerical parameters: the threshold value $p_5^P$ (analogous to $p_5^T$ for T), which determines one of the candidates for the P wave onset and offset is given in Table \ref{table2}.

The P wave boundaries refinement and the definition of the P wave morphology, follows the algorithm described in the \ref{delineation:t:morphology}, but with the numerical values of the parameters $p_6^P$ and $p_7^P$ changed for the P wave.

The refinement of the P wave, using the delineation results from all leads, is similar to the above algorithms for the QRS and T complexes. The corresponding numerical parameters $p_8^P$ and $p_9^P$ are also given in the table \ref{table2}.

\begin{table}[!ht]
	\begin{adjustwidth}{-2.25in}{0in} 
		\centering
		\caption{
			{\bf Parameters used in the delineation algorithm.}}
		\begin{tabular}{|c|c|c|}
			\hline
			&&\\[-1em]
			$p_1^{QRS}=0.5$ & $p_1^T=0.6$ & $p_1^P=0.4$ \\
			\hline
			&&\\[-1em]
			$p_2^{QRS}=0.05$ & $p_2^T=0.2$ & $p_2^P=0.08$ \\
			\hline
			&&\\[-1em]
			$p_3^{QRS}=0.3$ & $p_3^T=0.85$ & $p_3^P=0.9$ \\
			\hline
			&&\\[-1em]
			$p_4^{QRS}=0.2$ & $p_4^T=0.15$ & $p_4^P=0.05$ \\
			\hline
			&&\\[-1em]
			$p_5^{QRS}=0.075$ & $p_5^T=0.125$ & $p_5^P=0.1$ \\
			\hline
			&&\\[-1em]
			$p_6^{QRS}=0.05$ & $p_6^T=0.2$ & $p_6^P=0.2$ \\
			\hline
			&&\\[-1em]
			$p_7^{QRS}=0.15$ & $p_7^T=0.3$ & $p_7^P=0.15$ \\
			\hline
			&&\\[-1em]
			$p_8^{QRS}=0.75$ & $p_8^T=0.8$ & $p_8^P=0.7$ \\
			\hline
			&&\\[-1em]
			$p_9^{QRS}=2.0$ & $p_9^T=2.5$ & $p_9^P=2.0$ \\
			\hline
		\end{tabular}
		\label{table2}
	\end{adjustwidth}
\end{table}

\section{Algorithm results}\label{results}

To assess the quality of the proposed algorithm, we perform a computational experiment to compare automatically done delineation with delineation performed manually by a cardiologist. We used the open access QT database \cite{Laguna1997, Goldberger2000}. This database contains $105$ $15$-minute $2$-channel ECG records with a sampling frequency of $250$~Hz. For each record, cardiologists annotated manually from $30$ to $70$ beats (including P waves, QRS complexes and T waves).

Comparison of the results of the proposed algorithm with the QTDB database is performed by searching of the complex, marked by the cardiologist, for each complex, found by the algorithm, in the window of $150$~ms. The value of this interval is selected in accordance with the standard of the Association for the Advancement of Medical Instrumentation \cite{standart1999} that describes the metrics for assessing the quality of delineation algorithms. This time interval will be named as the \textit{acceptable time window} or \textit{tolerance}.

If the algorithm detects the point correctly, the correct result (\textit{true positive -- TP}) is counted, and the error value is calculated as the time interval between the manually annotated point from the database and the automatic detection, performed by the algorithm. The algorithm is executed separately for each lead of the QT database, and for each point the lead with the lowest error is selected. If there is no corresponding point in the database for the point, detected by the algorithm, a type I error (\textit{false positive -- FP}) is counted. If the algorithm does not detect a point in the database, a type II error (\textit{false negative -- FN}) is counted. Following \cite{Rincon2011, DiMarco2011, Martinez2004, Bote2017}, we define $4$ basic quality metrics for the delineation algorithm:

\begin{itemize}
	\item average error $m$;
	\item standard deviation of the mean error $\sigma$;
	\item sensitivity $Se(\%) = TP / (TP + FN)$;
	\item positive predictive value $PPV(\%) = TP / (TP + FP)$.
\end{itemize}
Here, $TP$, $FP$, $FN$ denotes the total number of correct solutions, type I errors, and type II errors, respectively.

Table~\ref{table3} presents comparative results for the QT database, obtained with the proposed algorithm, as well as those presented in \cite{Rincon2011, DiMarco2011, Martinez2004, Bote2017}. It should be noted that the size of the acceptable time window differs in the presented works, so the result of the comparison is not absolutely strict. In \cite{DiMarco2011} the length of the time window is $150$~ms, in~\cite{Bote2017}  $200$~ms, in \cite{Rincon2011} $320$~ms, and in \cite{Martinez2004} this information is not reported. The larger the value of this acceptable time window, the higher the sensitivity, positive predictive value and standard deviation. For the algorithm presented in this paper the value of the acceptable time window is selected in accordance with the standard \cite{standart1999} and equals to $150$~ms.

\begin{table}[h!!]
	\caption{Comparison of the delineation results of the algorithm, presented in this paper, with the results of other works}
	\label{table3}
	\begin{center}
		\resizebox{\textwidth}{!}{%
			\begin{tabular}{|c|c|c|c|c|c|c|c|c|}
				\hline
				\multicolumn{2}{|c|}{     }&&&&&&&\\[-1em]
				\multicolumn{2}{|c|}{     } & P onset & P peak & P offset & QRS onset & QRS offset & T peak & T offset \\
				\hline
				&&&&&&&&\\[-1em]
				Current work & \begin{tabular}{@{}c@{}}$Se (\%)$ \\ $PPV (\%)$ \\ $m\pm\sigma(ms)$ \end{tabular} & \begin{tabular}{@{}c@{}}$97.46$ \\ $97.86$ \\ $-3.5\pm13.8$ \end{tabular} & \begin{tabular}{@{}c@{}}$97.50$ \\ $97.89$ \\ $4.3\pm10.0$ \end{tabular} & \begin{tabular}{@{}c@{}}$97.53$ \\ $97.93$ \\ $3.4\pm12.7$ \end{tabular} & \begin{tabular}{@{}c@{}}$98.42$ \\ $98.24$ \\ $-5.1\pm6.6$ \end{tabular} & \begin{tabular}{@{}c@{}}$98.42$ \\ $98.24$ \\ $4.7\pm9.5$ \end{tabular} & \begin{tabular}{@{}c@{}}$98.24$ \\ $98.24$ \\ $7.2\pm13.0$ \end{tabular} & \begin{tabular}{@{}c@{}}$96.16$ \\ $94.87$ \\ $13.4\pm18.5$ \end{tabular} \\
				\hline
				&&&&&&&&\\[-1em]
				\begin{tabular}{@{}c@{}} Bote \textit{et al.} \cite{Bote2017} \\ Standard mode \end{tabular} & \begin{tabular}{@{}c@{}}$Se (\%)$ \\ $PPV (\%)$ \\ $m\pm\sigma(ms)$ \end{tabular} & \begin{tabular}{@{}c@{}}$98.12$ \\ $94.26$ \\ $23.9\pm19.5$ \end{tabular} & \begin{tabular}{@{}c@{}}$99.15$ \\ $95.11$ \\ $13.8\pm8.8$ \end{tabular} & \begin{tabular}{@{}c@{}}$99.87$ \\ $96.03$ \\ $-1.9\pm10.4$ \end{tabular} & \begin{tabular}{@{}c@{}}$99.50$ \\ $99.78$ \\ $6.4\pm5.5$ \end{tabular} & \begin{tabular}{@{}c@{}}$99.50$ \\ $99.78$ \\ $-5.2\pm10.8$ \end{tabular} & \begin{tabular}{@{}c@{}}$99.41$ \\ $98.96$ \\ $9.0\pm15.4$ \end{tabular} & \begin{tabular}{@{}c@{}}$96.98$ \\ $95.98$ \\ $-12.9\pm18.6$ \end{tabular} \\
				\hline
				&&&&&&&&\\[-1em]
				DiMarco \textit{et al.} \cite{DiMarco2011} & \begin{tabular}{@{}c@{}}$Se (\%)$ \\ $PPV (\%)$ \\ $m\pm\sigma(ms)$ \end{tabular} & \begin{tabular}{@{}c@{}}$98.15$ \\ $91.00$ \\ $-4.5\pm13.4$ \end{tabular} & \begin{tabular}{@{}c@{}}$98.15$ \\ $91.00$ \\ $-4.7\pm9.7$ \end{tabular} & \begin{tabular}{@{}c@{}}$98.15$ \\ $91.00$ \\ $-2.5\pm13.0$ \end{tabular} & \begin{tabular}{@{}c@{}}$100.00$ \\ $-$ \\ $5.1\pm7.2$ \end{tabular} & \begin{tabular}{@{}c@{}}$100.00$ \\ $-$ \\ $0.9\pm8.7$ \end{tabular} & \begin{tabular}{@{}c@{}}$99.72$ \\ $97.76$ \\ $-0.3\pm12.8$ \end{tabular} & \begin{tabular}{@{}c@{}}$99.77$ \\ $97.76$ \\ $1.3\pm18.6$ \end{tabular} \\
				\hline
				&&&&&&&&\\[-1em]
				Martinez \textit{et al.} \cite{Martinez2004} & \begin{tabular}{@{}c@{}}$Se (\%)$ \\ $PPV (\%)$ \\ $m\pm\sigma(ms)$ \end{tabular} & \begin{tabular}{@{}c@{}}$98.87$ \\ $91.03$ \\ $2.0\pm14.8$ \end{tabular} & \begin{tabular}{@{}c@{}}$98.87$ \\ $91.03$ \\ $3.6\pm13.2$ \end{tabular} & \begin{tabular}{@{}c@{}}$98.75$ \\ $91.03$ \\ $1.9\pm12.8$ \end{tabular} & \begin{tabular}{@{}c@{}}$99.97$ \\ $-$ \\ $4.6\pm7.7$ \end{tabular} & \begin{tabular}{@{}c@{}}$99.97$ \\ $-$ \\ $0.8\pm8.7$ \end{tabular} & \begin{tabular}{@{}c@{}}$99.97$ \\ $97.79$ \\ $0.2\pm13.9$ \end{tabular} & \begin{tabular}{@{}c@{}}$99.77$ \\ $97.79$ \\ $-1.6\pm18.1$ \end{tabular} \\
				\hline
				&&&&&&&&\\[-1em]
				Rincon \textit{et al.} \cite{Rincon2011} & \begin{tabular}{@{}c@{}}$Se (\%)$ \\ $PPV (\%)$ \\ $m\pm\sigma(ms)$ \end{tabular} & \begin{tabular}{@{}c@{}}$99.87$ \\ $91.98$ \\ $8.6\pm11.2$ \end{tabular} & \begin{tabular}{@{}c@{}}$99.87$ \\ $92.46$ \\ $10.1\pm8.9$ \end{tabular} & \begin{tabular}{@{}c@{}}$99.91$ \\ $91.70$ \\ $0.9\pm10.1$ \end{tabular} & \begin{tabular}{@{}c@{}}$99.97$ \\ $98.61$ \\ $3.4\pm7.0$ \end{tabular} & \begin{tabular}{@{}c@{}}$99.97$ \\ $98.72$ \\ $3.5\pm8.3$ \end{tabular} & \begin{tabular}{@{}c@{}}$99.97$ \\ $98.91$ \\ $3.7\pm13.0$ \end{tabular} & \begin{tabular}{@{}c@{}}$99.97$ \\ $98.50$ \\ $-2.4\pm16.9$ \end{tabular} \\
				\hline
				\multicolumn{2}{|c|}{     }&&&&&&&\\[-1em]
				\multicolumn{2}{|c|}{$2\sigma_{CSE}(ms)$} & $10.2$ & $-$ & $12.7$ & $6.5$ & $11.6$ & $-$ & $30.6$ \\
				\hline
		\end{tabular}}
	\end{center}
\end{table}

The results show that the sensitivity for the onsets and peaks of the P, QRS and T waves is higher than 97\%, and the standard deviation of the mean error for each marked point is below the tolerances of the Committee of General Standards for Electrocardiography (\textit{CSE}) \cite{standart1985} ($s<2 \sigma_{CSE}$), except for the P wave onset, for which this value is $3$~ms higher. The T wave offset has the biggest mean error among the presented ones, however, the correct segmentation of this point is a known problem even for cardiologists, which was discussed in \cite{Mehta2008}.

It can be seen that the sensitivity values ​​are very similar to the results obtained in other works, despite the fact that these values are directly related to the selected value of the acceptable time window. In particular, the positive predictive value PPV obtained in this paper is higher than for the other methods, for the P wave and lower for the T wave offset, which demonstrates the difficulty of delineating this point. For the standard deviation, which is the most important parameter illustrating the variance of errors, the results are consistent with the \textit{CSE} standard and show that the proposed algorithm can determine ECG reference points such as QRS onsets and offsets, or P peak and offsets, with a smaller dispersion than other methods.

\section{Conclusions}\label{conclusions}

A new algorithm is proposed for the determination of peaks, boundaries and other significant points of various waves of the ECG signal, such as P wave, QRS complex and T wave, taking into account information from all available leads and providing similar or higher accuracy in comparison with other modern technologies. To test the effectiveness of the proposed algorithm, the QT database was selected, and the results were compared with the annotations of cardiologists. The results show a sensitivity above 97\% when detecting ECG wave peaks and 96\% for onsets and offsets, as well as better positive predictive value compared with previous works. Moreover, delineation errors of all control points remain within the tolerances defined by the Committee of General Standards for Electrocardiography (CSE), with the exception of the P wave onsets, for which the algorithm exceeds the tolerance for a small fraction of the duration of the complex.

The presented algorithm has several main advantages. First, it is possible to form the complete morphology of each complex in the form of a list of points characterized by the name, index, and value of the potential. This information can be used by a physician to determine symptoms of cardiac arrhythmias and cardiovascular diseases. In addition, this approach allows us to analyze such complex phenomena as late ventricular and atrial potentials. Secondly, the refinement, which taking into account the set of leads allows improving the quality of delineation: it increases the sensitivity by restoring the missed complexes and reduces the mean error and standard deviation. Improving the results will be observed with an increase in the number of available ECG signal leads: for $12$ leads there will be more available information on the complexes than for the $2$ leads. The described advantages allow us to qualify the presented algorithm of ECG segmentation as a useful tool for assisting physicians in revealing symptoms of cardiovascular diseases and diagnosing.

The work was supported by the Ministry of Education of the Russian Federation (contract No. 02.G25.31.0157 of 01.12.2015).


\end{document}